\documentclass[prl,showpacs,nobibnotes,preprintnumbers,superscriptaddress,nofootinbib,floatfix,twocolumn]{revtex4-1}
\usepackage{amsmath,amssymb,amsbsy,color}
\usepackage{epsfig}
\usepackage{psfrag}
\usepackage{epstopdf}
\usepackage{grffile} 
\usepackage{multirow}
\usepackage{algorithm}
\usepackage{algorithmic}

\def\red{}
\def\Red{}
\def\blue{}
\def\Blue{}
\def\green{}
\def\cyan{}
\newcommand{\TODO}[1]{}
\newcommand{\IGNORE}[1]{}

\def\bal#1\eal{\begin{align}#1\end{align}} 
\def\bgat#1\egat{\begin{gather}#1\end{gather}} 

\newcommand{\calO}{{\cal O}}
\newcommand{\vev}[1]{\langle #1 \rangle}

\def\calOi{{\cal O}^{(\text{imp})}}
\def\calOa{{\cal O}^{(\text{appx})}}
\def\calOag{{\cal O}^{(\text{appx}),g}}
\def\calOaG{{\cal O}^{(\text{appx})}_G}
\def\calOr{{\cal O}^{(\text{rest})}}

\def\Nc{N_\text{conf}}

\def\1{\mathbbm{1}}

\newcommand{\NoWordCount}[1]{#1}

\usepackage{ifpdf}
\ifpdf 
\DeclareGraphicsExtensions{.pdf}
\usepackage{hyperref}
\hypersetup{
    colorlinks=true,
    linkcolor=black,
    citecolor=black,
    filecolor=black,
    urlcolor=black,
}
\else
\fi

\begin{document}
\preprint{RBRC-967}
\title{A new class of variance reduction techniques using lattice symmetries}

\author{Thomas Blum}
\affiliation{
  Physics Department, University of Connecticut, Storrs, CT 06269-3046, USA
}
\affiliation{
  RIKEN-BNL Research Center, Brookhaven National Laboratory, Upton, NY 11973, USA
}

\author{Taku Izubuchi}
\affiliation{
 Brookhaven National Laboratory, Upton, NY 11973, USA
}
\affiliation{
  RIKEN-BNL Research Center, Brookhaven National Laboratory, Upton, NY 11973, USA
}

\author{Eigo Shintani}
\affiliation{
  RIKEN-BNL Research Center, Brookhaven National Laboratory, Upton, NY 11973, USA
}

\begin{abstract}

We present a general class of unbiased improved estimators for physical observables in lattice gauge theory computations which significantly reduces statistical errors at modest computational cost.  
The error reduction techniques, referred to as covariant approximation 
averaging, utilize approximations which are covariant under lattice symmetry transformations. {\Red We observed cost reductions from the new method compared to the traditional one, for fixed statistical error, of
{\Blue16 times} for the nucleon mass at $M_\pi\sim 330$ MeV (Domain-Wall quark) and 
{\Blue 2.6-20 times} for the hadronic vacuum polarization at $M_\pi\sim 480$ MeV (Asqtad quark). These cost reductions should improve with decreasing quark mass and increasing lattice sizes.}

\end{abstract}
\NoWordCount{\maketitle}


As non-perturbative computations using lattice gauge theory are applied to a wider range of physically interesting observables, it is increasingly important to find numerical strategies that provide precise results. In Monte Carlo simulations our reach to important physics is still often limited by statistical uncertainties. 
Examples include hadronic contributions to the muon's anomalous magnetic moment~\cite{Aubin:2012me}, nucleon form factors and structure functions~\cite{Yamazaki:2009zq}, 
including nucleon electric dipole moments
\cite{Shintani:2005xg,Berruto:2005hg,Shintani:2006xr,Shintani:2008nt}, 
hadron matrix elements relevant to flavor physics ($e.g.$, $K\to\pi\pi$ amplitudes) \cite{Blum:2011ng}, 
and multi-hadron state physics~\cite{Savage:2011xk}, to name only a few.

As a generalization of low-mode 
averaging (LMA)~\cite{Giusti:2004yp,DeGrand:2004qw}, we present a class of unbiased statistical error reduction techniques, utilizing approximations that are covariant under lattice symmetry transformations.
LMA has worked well in cases where low eigenmodes of the Dirac operator dominate{\green~\cite{Neff:2001zr}}: low energy constants in the $\varepsilon$-regime
\cite{Giusti:2002sm,Giusti:2004yp,DeGrand:2005vb,Luscher:2007se,Fukaya:2007fb}, pseudoscalar meson masses and decay constants~\cite{Giusti:2005sx,Noaki:2008iy,Li:2010pw}, an so on.
With a modest increase in computational cost, the generalized method can reduce statistical errors by an order of magnitude, or more, even in cases where LMA fails.


Unlike LMA,
we account for all modes of the Dirac operator, averaging over (most of) the lattice volume,
with modest additional computational cost. 
{\Red The all-to-all methods {\Red \cite{Bali:2005fu,Foley:2005ac}} implement this stochastically for the higher modes, while treating the low-modes exactly.}
{\Blue For expectation values invariant under translations,}
statistics effectively increase by averaging over the whole lattice. The all-to-all method is advantageous when the stochastic noise introduced in the target observable
is comparable to, or smaller than, the gauge field fluctuations of the ensemble~\cite{Bali:2009hu}, which typically holds only for many random source vectors per measurement.
The error reduction techniques presented here, which do not rely on stochastic noise, 
{\Blue are potentially } more effective, provided an inexpensive approximation can be found for the desired observable.

In lattice gauge theory simulations an ensemble of
gauge field configurations $\{U_1, \cdots, U_{\Nc}\}$ is generated randomly, according to the Boltzmann weight, $e^{-{\cal S}[U]}$, where ${\cal S}[U]$ is the lattice-regularized action. 
The expectation value of a primary, covariant observable, $\calO$, 
\NoWordCount{
\begin{align}
\vev{\calO} = {1\over \Nc} \sum_{i=1}^{\Nc} \calO[U_i] + {O\left({1 \over \sqrt{\Nc}}\right)},
\end{align}
}
is estimated as the ensemble average, over a large number of configurations, $\Nc \sim{ O}(100-1000)$. 
Here, we primarily consider observables made of fermion propagators $S_F[U]$ computed 
on the background gauge configuration $U$. 

By exploiting lattice symmetry transformations $g\in G$, 
that transform $U\to U^g$, a general class of variance reduction {\cyan techniques} is introduced. First construct an {\it approximation} $\calOa$ to $\calO$ which must fulfill the following conditions,
\begin{description}
\item[appx-1] $\calOa$ should fluctuate closely with
$\calO$, \\
$r \equiv   \text{Corr}(\calO,\calOa ) =
{ \vev{\Delta\calO \Delta\calOa} \over \sqrt{\vev{(\Delta\calO)^2} 
    \vev{(\Delta\calOa)^2}}} \approx 1$,
and $\langle (\Delta \calO)^2\rangle\approx\langle (\Delta \calO^{\rm (appx)})^2\rangle$
{, where $\Delta X = X - \vev{X}$.}
%
\item[appx-2] the cost to compute $\calOa$ is smaller 
than $\calO$'s,
$ \text{cost}(\calOa) \ll \text{cost}(\calO)$.
\item[appx-3] {\red $\vev{\calOa}$ is {\it covariant} under a lattice symmetry transformation, $g\in G$,
$  \vev{\calOa[U^g]} = \vev{\calOag[U]}$ {\cyan (in the examples below, a stronger condition holds: $\calOa$  is covariant on each configuration, rather than on average, $\calOa[U^g] = \calOag[U]$).}}
\end{description}
Note $\calOa$ and $\calOag$ refers to the approximations before and after
applying a  symmetry transformation $g$.

Using $\calO$ and $\calOa$ one can define an improved observable
\NoWordCount{
\begin{eqnarray}
\calOi &=& \calOr + \calOaG,
\label{eq:impObs}\\
\calOr &=& \calO - \calOa,
~\calOaG = {1\over N_G} \sum_{g\in G} \calOag,\nonumber
\end{eqnarray}
}
where an {average} over $N_G$ symmetry transformations in $G$ is taken.

For {\bf appx-1}, 
the statistical error of $\vev{\calOi}$ is
\NoWordCount{
\begin{eqnarray}
{\rm err}_{(\rm imp)}
&
\approx & {\text{err} \sqrt{2(1-r)+\frac{1}{N_G}}~~~,}
\label{eq:sigma sq}
\end{eqnarray}
}
which can be made smaller than the original (err) by a judicious choice of $\calOa$.
The fluctuation from $\calOr$,
the first term in (\ref{eq:sigma sq}), 
is suppressed due to $r\approx 1$, while 
the second term is reduced by $1/N_G$ 
without too much additional cost as required 
by {\bf appx-2} (correlations among $\calO, \calOa$, and $\calOag$ have been ignored, {\red which is a good approximation for noisy observables or large volume}). 
Due to covariance, {\bf appx-3},  it is easy to prove
the ensemble averages of (primary observables) 
$\calOa, \calOag$, and $\calOaG$ are all equal, 
so the improved estimator (\ref{eq:impObs}) is unbiased,
$\vev{\calOi} = \vev{\calO}.$ 

{\blue The idea of exploiting covariance
\cite{Giusti:2004yp,DeGrand:2004qw} to improve statistical errors has a wider range of applicability than LMA, so in general we call it {\it covariant approximation averaging} (CAA).}
Several comments on CAA follow. From Eq.~\ref{eq:sigma sq}
the accuracy of the approximation $\calOa \approx \calO$ ({\bf appx-1}) 
should be precise enough so that the statistical error from $\calOr$ is below,
say, one-half of the desired final precision.
Too accurate an approximation wastes resources.
In $\calOi$, most of the statistical fluctuation is carried by $\calOa$, 
which is reduced by averaging over $N_G( \gg1)$ measurements with 
smaller cost ({\bf appx-2}).
Balance between these opposing parts of the method allows CAA
to reduce statistical errors significantly while keeping 
the computational cost low.

In the framework of CAA the best choice of 
approximation depends on the target observables and lattice parameters 
such as quark mass and volume.
In principle, any set of lattice symmetries, $G$, can be used in CAA. 
We limit ourselves to the case of translation symmetries 
in the following examples.

The first example is LMA. 
In LMA eigen-systems of the Hermitian Dirac operator are obtained for the part 
of the spectrum closest to zero,
\NoWordCount{
\begin{align}
D_H v_i &= \lambda_i v_i, ~~ (i=1,2,\cdots, N_\text{eig}),\\
&0 < |\lambda_1| \leq |\lambda_2| \leq \cdots \leq |\lambda_{N_\text{eig}}| = \lambda_\text{cut},
\end{align}
}
which is then used to construct, through spectral decomposition, the low-mode approximation of the fermion propagator,
\NoWordCount{
\begin{align}
S_\text{LM}(x,y) = \sum_{i=0}^{N_\text{tot}} v_i(x) f_\text{LM}(\lambda_i)   
v_i^\dagger(y), \label{eq:SLMA}\\
f_\text{LM}(\lambda) = {1\over \lambda} \theta(\lambda_\text{cut} - |\lambda| ).
\label{eq:fLMA}
\end{align}
}
$N_\text{tot}$ is the total dimension of the Dirac matrix.
The recipe for LMA in terms of the CAA master Eq.~(\ref{eq:impObs}) is 
{\Red shown in left column of Table \ref{alg_LM_AM}.}
\NoWordCount{
\begingroup
\squeezetable
\begin{table}
\begin{center}
\caption{ LMA and AMA algorithms}
\label{alg_LM_AM}
{\Red
\begin{tabular}{ll|ll}
\hline
 & {\bf LMA algorithm} & & {\bf AMA algorithm} \\
\hline
1: & Compute low-modes $v_i$ of $D_H$ &
1: & if $\lambda_{\rm cut}\ne 0, N_\text{eig}>0$\\
& & & Compute low-mode $v_i$ of  $D_H$ \\\hline
2: & \multicolumn{3}{l}{Set source $b$ and $G-$invariant inital guess $x_0$}\\\hline
3: & \multicolumn{3}{l}{Compute exact $S$ and $\mathcal O[S]$ precisely (use deflation if $v_i$ exits)}\\\hline
4: & Repeat for $S_{\rm LM}$ in (\ref{eq:SLMA}) & 
4: & Repeat for $S_{\rm AM}$  in (\ref{eq:SAMA}) \\
   & and $\calOa = \calO[S_\text{LM}]$ &
   & and $\calOa = \calO[S_\text{AM}]$  using\\
& & & deflated CG (if $\lambda_{\rm cut}\ne 0$)\\\hline
5: & $\calOr=\calO[S]-\calO[S_\text{LM}]$&
5: & $\calOr=\calO[S]-\calO[S_\text{AM}]$; \\\hline
6: & \multicolumn{3}{l}{Set shifted source $b^{\Blue g}$ and  $G-$invariant inital guess $x_0^{\Blue g}$}\\\hline
7: & Average $\calOag=\calO[S_\text{LM}]$ & 
7: & Average $\calOag=\calO[S_\text{AM}]$ \\
   &   over $g \in G$  to get $\calOaG$ &
   &   over  $g \in G$ to get $\calOaG$  \\\hline
8:   & \multicolumn{3}{l}{ $\calOi = \calOr + \calOaG$}\\\hline
\end{tabular}
}
\end{center}
\end{table}
\endgroup
}
Although LMA is particularly good for observables dominated 
by low-modes, such as the single pion state for lighter fermion masses, 
LMA does not work so well for heavier hadrons or 
when the quark mass is heavier {\red\cite{Giusti:2005sx,Li:2010pw}}
{\green (see also \cite{Bali:2010se} for dependence on parity of states and (non-)Hermiticity of Dirac operators)}.
This is due to the truncation of the sum in (\ref{eq:SLMA}),  {\it i.e.}, 
$f_\text{LM}(\lambda) =0$ for $|\lambda| > \lambda_\text{cut}$.

One could improve the above 
by  constructing a {polynomial } for $1/\lambda$ 
and using it to obtain 
a better (all-mode) approximation of the propagator above $\lambda_\text{cut}$:
\NoWordCount{
\begin{align}
S_\text{AM}(x,y) = \sum_{i=0}^{N_\text{tot}} v_i(x) f_\text{AM}(\lambda_i)   
v_i^\dagger(y), \label{eq:SAMA}\\
f_\text{AM}(\lambda) = 
\begin{cases}
{1\over \lambda} & |\lambda| \leq \lambda_\text{cut}\\
P_n(\lambda)      & |\lambda| > \lambda_\text{cut}
\end{cases}
\label{eq:fAMA}
\end{align}
}
where $P_n(\lambda)\red \approx 1/\lambda$ is a polynomial of degree $n$,
From (\ref{eq:SAMA}) and (\ref{eq:fAMA}), one computes the approximate
propagator using $P_n(D_H)$ in the subspace orthogonal to  
the eigenvectors below $\lambda_\text{cut}$,
\NoWordCount{
\begin{align}
S_\text{AM} &= 
\sum_{i=1}^{N_\text{eig}} v_i{1\over \lambda_i} v_i^\dagger
+ P_n(D_H)  (1-\sum_{i=1}^{N_\text{eig}} v_i v_i^\dagger),
\end{align}
}
with number of low-modes $N_{\rm eig}$.
In analogy to LMA, we refer to the above as all-mode averaging (AMA). A recipe similar {\Red to LMA is shown in the right column of  Table \ref{alg_LM_AM}.}

{\blue As emphasized in~\cite{Giusti:2004yp} approximate eigenvectors can be used in LMA (and AMA) to reduce the cost of this part of the calculation. We have not done that as we find the cost of computing them exactly is not too burdensome and is partly recouped in the deflation of the Dirac operator. }

Among many different ways~{\Red\cite{{Luscher:1993xx,Borici:1995am,Kamleh:2011dc}}} to obtain $P_n(\lambda)$,
one of the easiest is to use the polynomial implicitly generated
by an iterative linear solver such as conjugate gradient (CG). 
For example (\ref{eq:SAMA}) can be implemented
as a CG solution using the low-mode approximation
applied to the source vector $b$
(the coefficients of $P_n$ depend on $b$)
as the starting vector, $S_\text{LM} b$, which is 
nothing but a deflated CG with iteration number set to the degree of the polynomial, $n$. 
One can either fix $n$ (number of iterations) or
the CG residual vector stopping criterion. Either
satisfies the covariance condition ({\bf appx-3}).  
{\green
This particular construction of $P_n(D_H)$ is called the
truncated solver method (TSM)~\cite{Bali:2009hu}.
The difference with AMA {\cyan is that TSM is applied in \cite{Bali:2009hu} to a random source,
and the unbiased result is guaranteed by stochasticity} 
while AMA relies on covariance, so it does not need the random source. 

In \cite{Li:2010pw} low-modes are  utilized with $Z_3$ noise to compute many-to-all hadron correlation functions for variance reduction.
}
{\green One may also choose $N_\text{eig}=0, \lambda_\text{cut}=0$ in (\ref{eq:fAMA}), $i.e.$ not to use eigenvectors at all.  This may be effective for heavier quark masses, {\cyan but for lighter quarks one needs a larger degree polynomial for an accurate approximation and  $N_\text{eig}>0$ is likely more cost-effective}. }

To ensure unbiasness one should check, on a few configurations, the covariance of the particular implementation of the approximation 
$\calOa[U^g] = \calOag[U]$  by computing the approximation explicitly on a transformed gauge field to compare with the original gauge field to see that they are equivalent to numerical precision. 

To compare the LMA and AMA methods, we use
the 2+1 flavor Domain-Wall fermion (DWF) ensemble generated by the
RBC/UKQCD collaboration \cite{Aoki:2010dy} with
lattice size $24^3\times 64$, extra dimension size $L_s=16$, and 
Iwasaki gauge action ($\beta=2.13$, or $a^{-1}=1.73$ GeV). 
The low-modes of the Hermitian DWF Dirac operator are obtained using
a 4D-even-odd-preconditioned, shifted Lanczos algorithm {\green\cite{Neff:2001zr}}
with accuracy $\|(D_H-\lambda_i)v_i\|/\|v_i\| < 10^{-12}$.
The eigen-modes are used for LMA as in Eqs.~(\ref{eq:SLMA}) 
and (\ref{eq:fLMA}), to deflate the CG, and to evaluate the low-mode parts of both $\mathcal O$ and
$\mathcal O^{(\rm appx)}$ {\Blue, and similarly} for AMA as in Eqs.~(\ref{eq:SAMA}) and (\ref{eq:fAMA}). 
In this paper we compute 180 low-modes for light quark mass $m=0.01$ and 400 low-modes
for $m=0.005$.

We adopt translational symmetry on the lattice as $G$ and take $N_G$ propagator source locations, starting from the origin, separated by 12 lattice units in space and 16 in time, and
the total set of translations numbers $N_G=2^3\times 4=32$.
For AMA, the stopping condition of the ``sloppy CG''
for our approximation is $\|D_Hx-b\|/\|b\|< 3\times 10^{-3}$
{\Blue while it is $10^{-8}$ in \cite{Yamazaki:2009zq}.}
Note that when using an even-odd preconditioned Dirac operator, LMA and AMA guarantee unbiased estimators for translations by an even number of sites ({\bf appx-3}). We have explicitly checked this in our calculations.

%


Table \ref{tab:prop} lists the relative statistical errors for various hadronic two-point correlation functions computed using LMA, AMA, and the original CG method, for $m=0.005$.
All were obtained with the same Gaussian smeared sources and {\Red point (Gaussian) sinks for pseudoscalar and vector (Nucleon)} used in \cite{Yamazaki:2009zq}.
At short distance $(t=4)$, there is no improvement between the original and LMA cases, except in the pseudoscalar (PS) channel.
This is because the contribution of higher modes is still important in the short-distance region. 
{\red Although for LMA $N_G$ could be taken as large as the lattice size with modest cost,  
we set $N_G=32$ since {\cyan larger} $N_G$ is not effective due to correlations between nearby gauge fields {\cyan in our examples}.}
On the other hand, AMA dramatically reduces the errors (more than 4-6$\times$) for all channels 
(and different momenta) and for all distances.
In this example the variance reduction by AMA comes almost entirely from the second term in Eq.~(\ref{eq:sigma sq}) 
since $r{\Blue =\text{Corr}(\calO,\calOa)}$ is very close to one
 ($r>0.9999$ for $m=0.005$), even though the residual stopping criterion used for $\calOa$ is loose 
($3\times 10^{-3}$).
For LMA at short distance $r\simeq0.9$ so the error from $\calOr$ is significant.
We also confirm that for the PS channel both LMA and AMA yield improvement, with $r>0.997$ even in the short distance region, as suggested previously for LMA using overlap fermions 
\cite{Noaki:2008iy,Li:2010pw}. {\red For $m= 0.01$ $r$ is somewhat smaller ($r>0.99$), so the contribution from $\calOr$ is more significant. Only 180 low-modes were used for $m=0.01$. }

\begin{table}
\begin{center}
\caption{Correlation function relative statistical error for $N_{\rm conf}=109$ (separated by 40 trajectories) and $N_G=32$. Nucleon (N), pseudoscalar (PS), and vector (V) channels. $m=0.005$.
 Gaussian smeared sink is used for the nucleon, others are point sinks. 
\IGNORE{$n_p$ refers to units of momenta.}
{\Red Gaussian smeared source is used for all channels.}
}
\NoWordCount{
\label{tab:prop}
\begin{tabular}{ccccc}
\hline\hline
Hadron & $t$ & Original [\%] & LMA  [\%] &  AMA [\%] \\
\hline
N 
&4 & 6.9 & 5.0 & 1.5  \\
\IGNORE{(${n_p^2=0}$)}
& 8 & 9.2 &3.2 & 1.9  \\
& 12 & 23 & 4.8  & 3.5 \\\hline 
\IGNORE{
N
&4& 6.8 & 5.1  & 1.5  \\
(${n_p^2=1}$) 
&8& 9.5 & 4.0  & 2.0 \\
&12& 18 & 4.6  & 3.3  \\\hline 
}
PS 
&4& 4.5  & 0.98 & 0.86 \\
&12& 4.9 & 0.91 & 0.86  \\
&28& 5.0 & 1.3  & 1.3  \\\hline 
V  
&4& 3.9 & 2.9  & 0.6  \\
&8& 5.2 & 2.1  & 1.1  \\
&12& 12 & 3.4 & 2.3  \\
\hline\hline
\end{tabular}
}
\end{center}
\end{table}

\begin{figure}[tb]
\begin{center}
  \vskip 2mm
  \includegraphics[width=85mm]{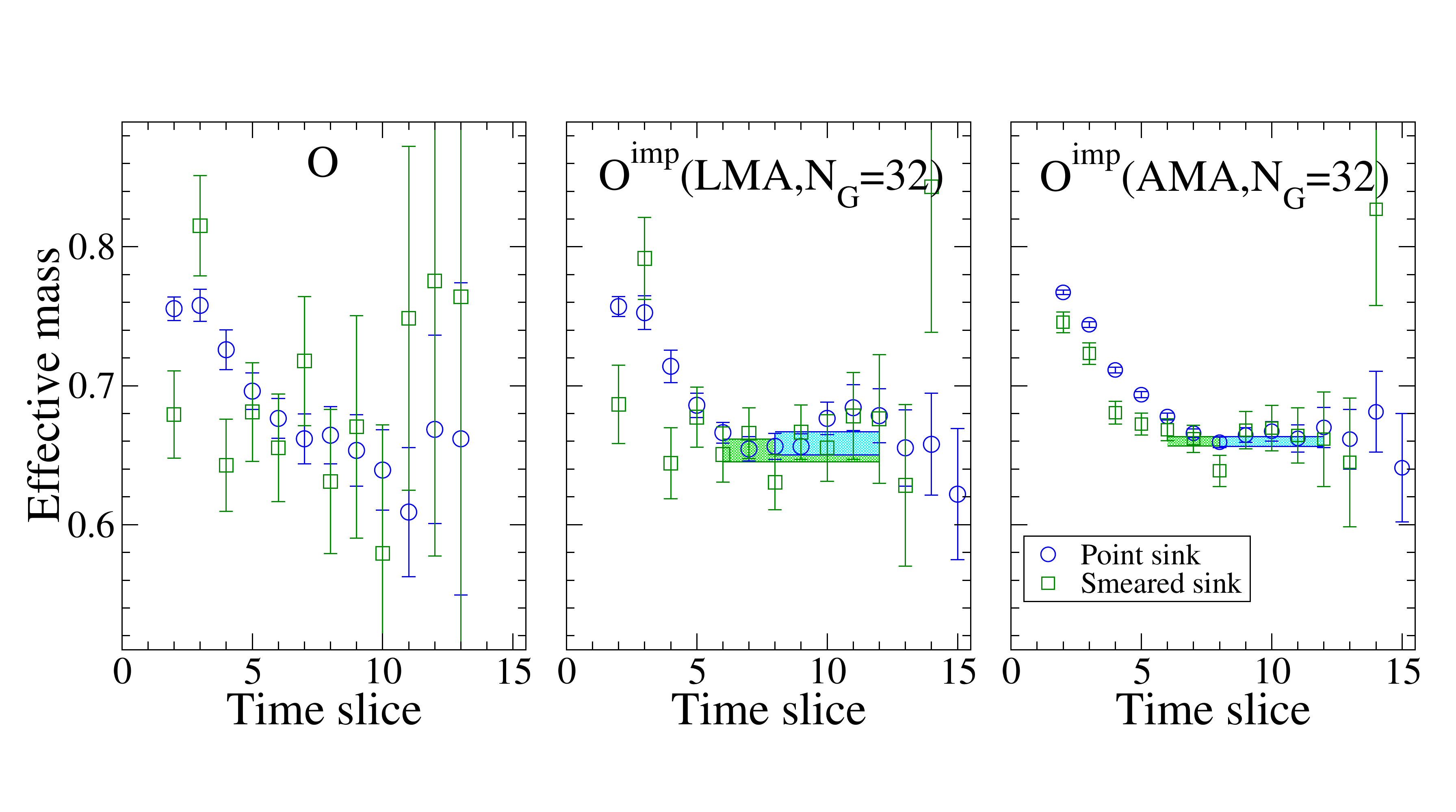}
  \caption{Nucleon effective mass using LMA (middle) and AMA (right). 
  $m=0.005$. Unimproved calculation (left).
 {\Red See Tab.~\ref{tab:prop} for parameters. }
  Colored bands denote fit mass and range. Gaussian sink.
  }
  \label{fig:effm}
\end{center}
\end{figure}

Figure \ref{fig:effm} shows the nucleon effective mass using LMA and AMA for the data in Tab.~\ref{tab:prop}, and Tab.~\ref{tab:mass} compares these to an earlier high statistics study of nucleon structure functions 
\cite{Yamazaki:2009zq}. The right-most panel in Fig.~\ref{fig:effm} shows significant improvement of  the effective mass plateau for AMA. 
Using the same fitting range, the precision of the nucleon mass attained with AMA is
smaller by more than a factor of 1.5 compared to the high statistics study \cite{Yamazaki:2009zq} 
where 3728 and 1424 measurements were made for $m=0.005$ and 0.01, respectively. {\Red The improved statistics make it easier to choose the fit range based on $\chi^2$, as seen in Fig.~\ref{fig:effm}}.
{\red LMA for nucleon masses was examined in \cite{Giusti:2005sx}.}

\begingroup
\begin{table}
\begin{center}
\caption{
Nucleon masses (GeV) using LMA, AMA
and from data from a high statistics study \cite{Yamazaki:2009zq}.
{\red See Table~\ref{tab:cost} for costs. Gaussian (gauss) and point (pt) sinks.}}
\NoWordCount{
\label{tab:mass}
\begin{tabular}{ccc|cc|c}
\hline\hline
 & & & \multicolumn{2}{c|}{$\calOi$, $N_G=32$} & \multicolumn{1}{c}{ $\calO$ } \\
  \hline
$m$  & sink & fit range & LMA & AMA &  High stat. \\
\hline
0.005 & pt & 8-12 &1.1391(145) & 1.1413(61) & 1.1561(104) \\
0.005 & gauss & 6-12 &1.1305(143) & 1.1420(58) & 1.1481(100)  \\
0.01  & pt & 9-15  & 1.2446(164)  & 1.2363(59) & 1.2101(89) \\
0.01  & gauss & 7-15 & 1.2240(148)  & 1.2268(60)  & 1.2169(93)  \\
\hline\hline
\end{tabular}}
\end{center}
\end{table}
\endgroup

Most of the cost of AMA comes from the low-mode and
sloppy CG parts of the approximation $\mathcal O^{(\rm appx)}$ (deflating the Dirac operator significantly reduces the cost of computing $\calOr$), and the larger $N_G$, the lesser the {\Blue relative} cost of the former.
The various costs for AMA in our examples are broken down in Table~\ref{tab:cost} and compared to the high statistics study~\cite{Yamazaki:2009zq}. 
In the example using Gaussian sinks,
AMA is roughly {\Blue 16 and 5 times} less expensive for roughly the same statistical error, for $m=0.005$ and 0.01, respectively. {\Red LMA is significantly less effective, {\Blue 3.6} and 2.3 times less expensive.
As $N_G$ increases, AMA improves statistics with relatively little extra cost. 
For instance, for $N_G=64$ AMA costs an additional {\Blue 114}, in units of the original propagator. 
The advantage of AMA clearly grows with increasing lattice size and decreasing quark mass. 
{\Red  The cost of calculating the correlation functions in this example is negligible, but this may not be the case for more complicated observables. Although disk space and CPU time for eigenvector
I/O can be non-negligible, we ignore these as the costs strongly depend
on the implementation details ($e.g.$, we could (de)compress
eigenvectors) and the features of the I/O systems used.}

Another impressive example of AMA is shown in Fig.~\ref{fig:hvp}, which depicts the hadronic vacuum polarization (HVP) from~\cite{Aubin:2012me} and using AMA for roughly the same amount of computational resource (20 configurations, 1400 low-modes {\red with accuracy $\|(D_H-\lambda_i)v_i\|/\|v_i\| < 10^{-10}$},
$N_G=708$, and {\Blue sloppy} CG stopping residual criterion $10^{-4}$  compared to $10^{-8}$ in \cite{Aubin:2012me}). The pion mass is $m_\pi=476$ MeV {\Red and lattice size $48^3\times144$}. The HVP contribution to the muon's anomalous magnetic moment is sensitive to the low $Q^2$ region~\cite{Aubin:2012me}, so constraining the HVP in this region is crucial to precisely extract the anomaly. In this test case (which was not optimized), {\Blue to achieve the same errors on the HVP in the range 0-1 GeV$^2$ as the original calculation required about 2.6-20 times less computer time.}
Interestingly, LMA actually increases the error in this case by about $2-3\times$ because the low-modes do not saturate the Ward-Takahashi identity. The stopping criterion for $\calOa$ can not be too low for the same reason, though our choice may have been too conservative. {\Red The costs are summarized in Tab.~\ref{tab:cost}. We note that in this case the cost of constructing the low mode part of the propagator is roughly equivalent to the sloppy CG cost, and that here again the contraction costs are negligible.}
\begin{figure}[tb]
\begin{center}
  \includegraphics[width=0.85\columnwidth]{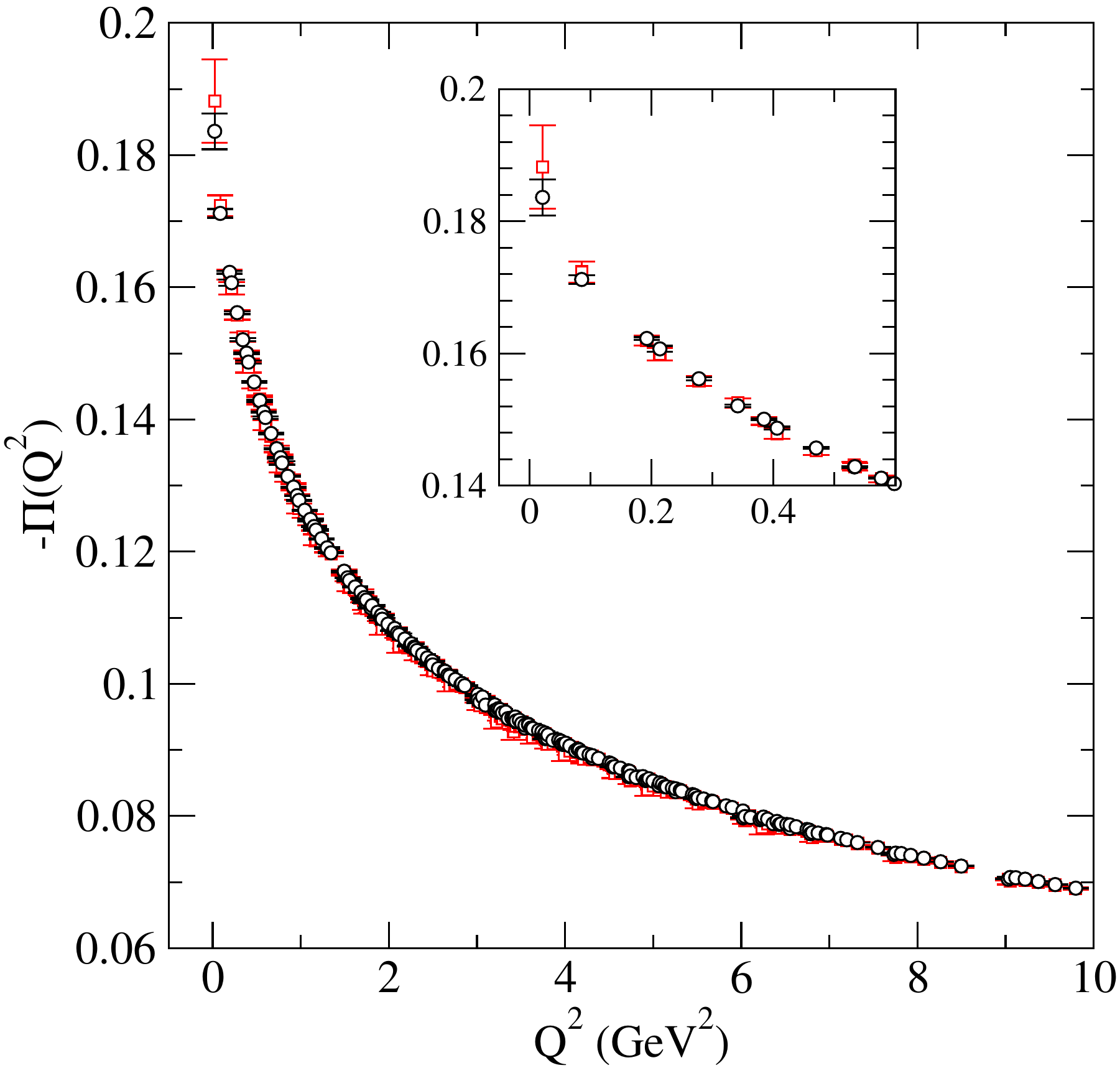}
  \caption{Hadronic vacuum polarization from~\cite{Aubin:2012me} (squares) and using AMA (circles). AMA {\Blue achieves the same statistical error as the original calculation in the range 0-1 GeV$^2$ for about 2.6-20 times less computer time. See Tab.~\ref{tab:cost} for details.}}
  \label{fig:hvp}
\end{center}
\end{figure}

\begingroup
\begin{table}
\begin{center}
\caption{Computational cost.
The unit of cost is one quark propagator without deflated CG, per configuration.
$N_G=32$ for nucleon masses and 708 for HVP. The last column gives the cost to achieve the same error for each method, {\Red normalized to~\cite{Yamazaki:2009zq} (nucleon mass $m_{N}$) and \cite{Aubin:2012me} (HVP)} and scaled by the errors in Tab.~\ref{tab:mass}. {\Red HVP scaled costs are maximum and minimum in the range $Q^2=0-1$ GeV$^2$. For $m=0.005$, in~\cite{Yamazaki:2009zq}, non-relativistic spinors were used which means the scaled costs in this case {\Blue were} increased by two. The cost of $\mathcal O_G^{\rm (appx)}$ for AMA is split to show the sloppy CG and low-mode costs separately.} }
\NoWordCount{
\label{tab:cost}
\begin{tabular}{c|cc|cccccc}
\hline\hline
& $N_{\rm conf}$& $N_{\rm meas}$ & LM & $\mathcal O$ &$\mathcal O_G^{\rm (appx)}$ & Tot. &  \multicolumn{2}{c}{scaled cost}\\ \hline
$m_N$ &
\multicolumn{6}{c}{\red $m=0.005$, 400 LM} & gauss & pt\\
\hline
AMA & 110  & 1 & 213 & 18 & 91+23 & 350 & 0.063 & 0.065 \\
{LMA} & 110 & 1 & 213 &  18 & 23 & 254 & 0.279 & 0.265\\
Ref. \cite{Yamazaki:2009zq}  & 932 & 4 & - & 3728 & - & 3728\footnote{
In~\cite{Yamazaki:2009zq} a doubled source was used to reduce this cost by two.} & 1 & 1\\
\hline
\multicolumn{8}{c}{\red $m=0.01$, 180 LM}\\
\hline
AMA        & 158 & 1  & 297 & 74  & 300+22 & 693 & 0.203 & 0.214  \\
{LMA} & 158 & 1 & 297 & 74 & 22   & 393 & 0.699 & 0.937  \\
Ref. \cite{Yamazaki:2009zq}   & 356 & 4  & -   & 1424 & -   & 1424 & 1 & 1 \\
\hline 
HVP &
\multicolumn{6}{c}{\red $m=0.0036$, 1400 LM} & max & min\\
\hline
AMA        & 20 & 1  & 96 & 11   & 504+420 & 1031 & 0.387 & 0.050 \\
{LMA} & 20 & 1 & 96 & 11 &  420  & 527 & 10.3 & 3.56 \\
Ref. \cite{Aubin:2012me}   & 292 & 2  & -   & 584 & -   & 584 & 1 & 1 \\
\hline\hline
\end{tabular}
}
\end{center}
\end{table}
\endgroup
{\cyan In this letter a} new class of unbiased error reduction techniques is introduced, 
using approximations that are covariant under lattice symmetries.
This is a generalization of low-mode averaging
which reduces the statistical error for observables that are not dominated by low-modes. 
We have shown through several numerical examples that all-mode averaging is
a powerful example of CAA, performing better than LMA and works well even in cases where LMA fails.
In the examples given here, AMA reduced the cost by factors up to 
{\Blue $\sim20$}, over conventional computations, and these factors will only increase for larger lattice sizes and smaller quark masses.
The method has great potential for investigations of difficult but important physics problems where statistical fluctuations still dominate the total uncertainty, like
the nucleon electric dipole moment or hadronic contributions to the muon anomalous magnetic moment.
Since CAA works without introducing any statistical bias (so long as condition {\bf appx-3} holds), there are many possibilities that also satisfy {\bf appx-1} and {\bf appx-2}: One can construct $\calOa$
using different lattice fermions and parameters 
(mass, $L_s$ (for DWF), boundary conditions and so on).
{\green {\Red$\vev{\calOaG}$} can be measured on a larger number of 
gauge configurations, which is potentially advantageous 
for observables dominated by gauge noise such 
as disconnected diagrams. One may also consider other types of approximations such as the 
hopping parameter expansion used in \cite{Bali:2009hu}, or approximations 
at the level of hadronic Green's functions.} 


\NoWordCount{
\begin{acknowledgments}
Numerical calculations were performed using the RICC at RIKEN and the Ds cluster at FNAL. 
We thank Sinya Aoki, Rudy Arthur, Gunnar Bali, Peter Boyle, Norman Christ, Thomas DeGrand, Leonardo	Giusti, Shoji Hashimoto, Tomomi Ishikawa, Chulwoo Jung, Takashi Kaneko, Christoph Lehner, Meifeng Lin, Stefan Schaefer, Ruth Van de Water, Oliver Witzel, {\cyan Takeshi Yamazaki}, 
Jianglei Yu {\Blue  and other members of JLQCD,RBC,UKQCD collaborations} for valuable discussions and comments. 
{\Blue CPS QCD library \cite{CPS} and other softwares (QMP,QIO) are used, which are supported by USQCD and USDOE SciDAC program. }
This work was supported by the Japanese Ministry of Education Grant-in-Aid, Nos. 22540301 (TI), 23105714 (ES), 23105715 (TI) and U.S. DOE grants DE-AC02-98CH10886 (TI) and DE-FG02-92ER40716 (TB).
We also thank BNL, the RIKEN BNL Research Center, and USQCD for providing resources necessary for completion of this work.
\end{acknowledgments}

\bibliography{ref.bib}
}

\end{document}